# Thermodynamic Origin of Reaction Non-Uniformity in Battery Porous Electrodes and Its Mitigation

Fan Wang[1] and Ming Tang[1*]

1. Department of Materials Science & NanoEngineering, Rice University, Houston, TX 77005, USA.

* Corresponding author email: mingtang@rice.edu

**Abstract**

The development of non-uniform reaction current distribution within porous electrodes is a ubiquitous phenomenon during battery charging / discharging and frequently controls the rate performance of battery cells. Reaction inhomogeneity in porous electrodes is usually attributed to the kinetic limitation of mass transport within the electrolyte and/or solid electrode phase. In this work, however, we reveal that it is also strongly influenced by the intrinsic thermodynamic behavior of electrode materials, specifically the dependence of the equilibrium potential on the state of charge: electrode reaction becomes increasingly non-uniform when the slope of the equilibrium potential curve is reduced. We employ numerical simulation and equivalent circuit model to elucidate such a correlation and show that the degree of reaction inhomogeneity and the resultant discharge capacity can be predicted by a dimensionless reaction uniformity number. For electrode materials that have equilibrium potentials insensitive to the state of charge and exhibit significant reaction non-uniformity, we demonstrate several approaches to spatially homogenizing the reaction current inside porous electrodes, including matching the electronic and ionic resistances, introducing graded electronic conductivity and reducing the surface reaction kinetics.

**Introduction**

Electrodes in rechargeable batteries usually take the form of porous electrodes, which consist of a porous matrix of active materials and additives with the void space filled by an electrolyte. Porous cathodes and anodes are typically prepared in the layer format and sandwiched between separators and current collectors to form a single battery cell. During charge / discharge, ions migrating within the electrolyte undergo redox reaction(s) with electrons transported in the solid matrix at the pore wall surface, which is represented by $Li^+ + e^- \rightleftharpoons Li$ for lithium-ion batteries. The reaction flux is often described by the Butler-Volmer equation:

$$j_{in} = i_0 \left[\exp\left(\frac{\alpha F\eta}{RT}\right) - \exp\left(-\frac{(1-\alpha)F\eta}{RT}\right)\right] \qquad 1)$$

where the surface overpotential $\eta = \Phi_L - \Phi_S + U_{eq}$, $\Phi_L$ and $\Phi_S$ are the electrical potentials of the electrolyte and solid matrix, respectively, and $U_{eq}$ is the equilibrium or open-circuit potential of the active material.

Upon (dis)charging, $j_{in}$ is usually spatially inhomogeneous within porous electrodes especially in the depth direction of the electrode layer due to the ionic and electronic resistances of electrolytic and solid matrix phases. Such reaction non-uniformity limits the power output and is a main cause for the under-utilization of battery capacity at high rates. With the ever-growing need for higher energy density and the improvement in battery fabrication processes, the use of thick electrodes in Li-ion batteries has attracted increasing interest in recent years. However, reaction inhomogeneity becomes more severe with increasing electrode thickness, which leads to the inferior rate performance of thick electrodes and presents a major barrier to their commercial applications. A rational understanding of the origins of the non-uniform reaction distribution is thus critical for the design and optimization of battery systems at the cell level.

The importance of reaction non-uniformity to electrode performance has long been recognized since the early study of porous electrodes(1-4). Newman and Tobias theoretically examined the reaction current distribution in porous electrodes by deriving analytical solutions to the one-dimensional porous electrode model(4). They show that the reaction distribution is controlled by two dimensionless numbers

$$\delta = \frac{(1-\alpha)IL_{cat}nF}{RT}\left(\frac{1}{\kappa_{eff}} + \frac{1}{\sigma_{eff}}\right) \qquad 2)$$

and $\sigma_{eff}/\kappa_{eff}$ in the Tafel region (i.e. large overpotentials) of the charge transfer kinetics. Eq. 2 is expressed in terms of anodic reactions and can be applied to cathodic reactions by changing $1 - \alpha$ to $\alpha$. In the linear region (i.e. low overpotentials), $\delta$ is replaced another dimensionless number

$$\nu = \sqrt{\frac{nFai_0L_{cat}^2}{RT}\left(\frac{1}{\kappa_{eff}} + \frac{1}{\sigma_{eff}}\right)} \qquad 3)$$

Reaction non-uniformity intensifies at large $\delta$ or $\nu$, i.e. when the system has low effective ionic ($\kappa_{eff}$) and electronic ($\sigma_{eff}$) conductivities, large current ($I$) and/or exchange current density ($i_0$). On the other hand, the ratio between the electronic and ionic conductivities $\sigma_{eff}/\kappa_{eff}$ controls the symmetry of the reaction distribution. Reaction occurs preferentially near the separator (or current collector) when $\sigma_{eff}/\kappa_{eff} \gg 1$ (or $\sigma_{eff}/\kappa_{eff} \approx 0$), and develops on both sides of the electrodes when $\sigma_{eff}/\kappa_{eff} \approx 1$.

In their analysis, Newman and Tobias treated the open-circuit potential $U_{eq}$ of electrodes in Eq. 1 as a constant. For electrode materials used in Li-ion batteries, however, it is common that $U_{eq}$ depends on the extent of reaction or state of charge (SOC). A spatial gradient of $U_{eq}$ will therefore result from the inhomogeneous reaction flux within porous electrodes, which will reversely influence the reaction distribution through the contribution of $U_{eq}$ to $j_{in}$ in Eq. 1. In a recent study(5), we show that the reaction behavior of Li-ion porous electrodes is strongly affected by the SOC dependence of $U_{eq}$. When $U_{eq}$ varies significantly with SOC, which is typical of compounds exhibiting solid-solution behavior upon (dis)charging such as Li(NiMnCo)O$_2$ (NMC) and Li(NiCoAl)O$_2$ (NCA), electrode particles tend to have a uniform reaction rate within the salt penetration region during discharge. We refer to this type of electrode materials as *uniform-reaction* or UR-type electrodes. In contrast, many battery compounds such as LiFePO$_4$ (LFP) and Li$_4$Ti$_5$O$_{12}$ (LTO) undergo prominent first-order phase transitions upon Li composition swing and have wide voltage plateaus on their (dis)charge curves. For this group of electrodes whose $U_{eq}$ is insensitive to SOC, reaction flux is instead confined to a narrow reaction zone, which propagates within the porous electrodes as (dis)charge continues. We refer to this type of compounds as *moving-zone-reaction* or MZR-type electrodes. It is discovered that UR-type electrodes can deliver 1.7 – 2 times of the capacity utilization by MZR-type electrodes under otherwise same discharging conditions (electrode thickness and porosity, C rate, etc) (5). While reaction inhomogeneity in porous electrodes is usually considered a kinetic phenomenon, our study clearly reveals the important role of intrinsic thermodynamic properties of electrode materials, in particular the SOC dependence of $U_{eq}$.

In this work, we further the study on the thermodynamic origin of reaction non-uniformity in porous electrodes by considering a continuous spectrum of reaction behavior (UR, MZR and their intermediates) modulated by the $U_{eq}$ – SOC relation. The effect of the average slope of the $U_{eq}$(SOC) curve, or $\Delta U_{eq}$, on the reaction distribution and rate capability is examined by both pseudo-two-dimensional (P2D) porous electrode simulation(6-10) and an equivalent circuit model. A dimensionless "reaction uniformity" number $\lambda$ containing $\Delta U_{eq}$ is introduced to characterize the degree of reaction homogeneity. $\lambda$ provides quantitative predictions of the reaction zone width and discharge capacity of battery cells. Based on insights obtained from the analysis, we propose several approaches to improving the reaction uniformity of MZR-type electrodes and demonstrate their effectiveness in P2D simulations.

**Results and Discussions**

**I. P2D Simulations of Electrode Reaction Distribution**

The two distinct types of reaction behaviors, UR vs MZR, can be illustrated by the discharge process of NMC111 and LFP cathodes, respectively. Figure 1a-b displays the P2D simulation of an NMC111 half cell (i.e. with Li metal anode) discharged at 5mA/cm$^2$ in the electrolyte-tranport-limited regime. Details on the implementation of the P2D model are described in Appendix A. The NMC111 cathode is 200 μm thick and other simulation parameters are listed in Table A1. Figure 1a shows that a large intercalation flux develops near separator at the beginning of discharge but is soon homogenized across the electrode before the depth of discharge (DoD) reaches 0.1 and then remains uniform until the end of discharge. Consequently, the entire electrode is uniformly lithiated and the SOC of electrode particles varies homogeneously within the porous electrode, see Figure 1b. Such UR behavior as schematized in Figure 1c is representative of electrode materials whose equilibrium potentials ($U_{eq}$) have a strong SOC dependence such as NMC and NCA.

The discharge process of an LFP half cell is shown in Figure 1d-e. While the LFP electrode has the same thickness as the NMC111 cathode and is discharged at the same current density of 5 mA/cm$^2$, its reaction flux is highly localized throughout the discharge process. Figure 1d shows that an intercalation flux peak first forms on the separator side and then migrates towards the current collector as discharging proceeds. The peak corresponds to a moving narrow reaction front, which separates a largely lithiated electrode region near the separator and an unreacted region near the current collector, see Figure 1e. Such MZR behavior is idealized by the schematic shown in Figure 1e. It is representative of electrode materials with SOC-independent $U_{eq}$, e.g. LFP and LTO that go through first-order phase transformation(s) upon (de)lithiation.

As the above simulations demonstrate, NMC111 and LFP half cells have very different reaction distributions during discharge, with the latter exhibiting much stronger inhomogeneity. Because the two cells have similar electrode thickness, porosity and conductivities, such difference results from the intrinsic properties of the two active materials and specifically, their different SOC dependence of $U_{eq}$. To unambiguously illustrate the effect of the $U_{eq}$(SOC) curve on reaction uniformity, here we consider a model active material whose $U_{eq}$ (unit: V) is given by

$$U_{eq} = -\frac{\Delta U_{eq}}{4} \ln\left(\frac{f_{Li}}{1-f_{Li}}\right) + 3 \qquad 4)$$

where $f_{Li}$ is the occupancy fraction of available Li sites ($f_{Li}$ = 1 − SOC) fraction. As shown in Figure 2a, parameter $\Delta U_{eq}$ is equal to the slope of $U_{eq}$ at SOC = 0.5, i.e. $\Delta U_{eq} = \left.\frac{dU_{eq}}{df_{Li}}\right|_{f_{Li}=0.5}$, and controls the steepness of the $U_{eq}$(SOC) curve. P2D simulations are performed with varied $\Delta U_{eq}$ while all the other system properties are kept unchanged. Figure 2b-d present the evolution of the scaled Li concentration $\tilde{c}_s \equiv c_s/c_{s,max}$ in electrode particles in a half cell discharged at 0.5C when $\Delta U_{eq}$ = 0.001, 0.01 and 1V, respectively. The electrode thickness $L_{cat}$ is 200 μm and other parameters are listed in Table A1. The electronic conductivity $\sigma_{eff}$ and surface reaction rate constant $k_0$ used in the simulations are sufficiently large so that the discharging process is kinetically limited by electrolyte transport. The comparison shown in Figure 2b-d reveals that Li intercalation becomes more homogeneous as $\Delta U_{eq}$ increases. Characteristics of MZR and UR are evident at $\Delta U_{eq}$ = 0.001 and 1 V, respectively. At the intermediate $\Delta U_{eq}$ (0.01 V), electrode reaction exhibits a combination of MZR and UR behaviors, where significant intercalation flux exists within a reaction zone of finite width. As illustrated in Figure 2c, we may define a *scaled reaction zone width* $W_{RZ}$ as the inverse of the slope of $\tilde{c}_s(\tilde{X})$ at $\tilde{c}_s = 0.5$ and SOC = 0.5: $W_{RZ} = d\tilde{X}/d\tilde{c}_s|_{\tilde{c}_s=0.5, SOC=0.5}$. Figure 2e shows that $W_{RZ}$, which provides a measure of the reaction uniformity, increases monotonically with $\Delta U_{eq}$ and should approach 0 and ∞ in the limiting MZR and UR cases, respectively. The effect of $\Delta U_{eq}$ on the reaction distribution has direct consequence on the rate performance of the electrodes. As shown in Figure 2f, increasing $\Delta U_{eq}$ from 0.001 to 1 V significantly improves the discharge performance at high rates, with a 74% increase in the normalized discharge capacity DoD$_f$ at 2C and a 103% increase at 5C.

The effect of the slope of the $U_{eq}$(SOC) curve on the reaction uniformity in porous electrodes can be qualitatively understood as follows. Reaction flux at the electrode particle surface is controlled by the overpotential $\eta = \Phi_L - \Phi_S + U_{eq}$, where $\Phi_L$ and $\Phi_S$ are the electrical potentials of the electrolyte and active material, respectively. When discharging starts, $U_{eq}$ is initially uniform across the porous electrode, but the presence of ionic / electronic resistances in electrolyte / solid phase generates spatial gradients in $\Phi_L$ and $\Phi_S$, which results in inhomogeneous $\eta$ and therefore $j_{in}$ according to the Butler-Volmer equation (Eq. 1). When $U_{eq}$ has a strong SOC dependence, non-uniform $j_{in}$ gives rise to an inhomogeneous spatial distribution of $U_{eq}$. Locations with higher $j_{in}$ see a larger decrease in $U_{eq}$, which in turn causes $\eta$ and $j_{in}$ to drop more significantly than locations receiving lower $j_{in}$. Therefore, the SOC-dependent $U_{eq}$ serves as a "rectifier" to the non-uniform reaction distribution: it reduces the spatial gradient of $j_{in}$ until a constant reaction flux is reached within the porous electrode. The larger is the slope of $U_{eq}$(SOC), the stronger is such rectifying effect and the faster the electrode can establish a uniform reaction during discharging. On the other hand, SOC-insensitive $U_{eq}$ such as in LFP and LTO is not able to compensate the spatial gradients of $\Phi_L$ and $\Phi_S$ to help homogenize the reaction flux. When discharging is electrolyte-transport-limited, reaction will first occur at the separator, where $\eta$ is the largest, and continue until electrode particles in the local region are fully intercalated, after which the reaction front will move away from the separator like a traveling wave.

In the next section, we analyze the dependence of reaction distribution on $U_{eq}$(SOC) in a more quantitative manner based on an equivalent circuit model.

## II. Equivalent Circuit Model for Reaction Uniformity Analysis

While P2D simulations provide detailed predictions of the reaction distribution within porous electrodes, its numerical nature makes it less straightforward to illuminate the general relation between the degree of reaction uniformity and various battery cell properties. As an alternative, we consider the discharge process in a considerably simplified circuit model, with the goal to derive a tractable expression to quantify the reaction uniformity in terms of the slope of $U_{eq}$(SOC) and other relevant parameters. Let $L_R$ be the dimension of the reaction zone in which the intercalation flux is non-zero during discharging. As illustrated in Figure 3b, the model represents this portion of the electrode with an equivalent circuit, which divides the zone into two regions (region I and II). For simplicity, electrode particles in each region is assumed to undergo reaction uniformly and have the same SOC, $\Phi_S$ and $\Phi_L$. We treat the active material as a generalized capacitor, whose characteristic voltage–charge relation is given by $U_{eq}$(SOC). Its internal resistance is neglected as solid diffusion is assumed to be facile. The electronic resistance of the solid phase and ionic resistance of the electrolyte are represented by two resistors connecting region I and II, $R_S = L_R/\sigma_{eff}$ and $R_L = L_R/\kappa_{eff}$, respectively.

Assuming small surface overpotential $\eta$, we use the linearized Butler-Volmer equation to express the reaction current in region I and II:

$$I_k = \frac{FaL_R i_0}{2RT} \eta_k \qquad k = 1,2 \qquad \qquad 5)$$

where $a$ is the volumetric surface area of the electrode particles and the exchange current density $i_0$ is taken as a constant. Accordingly, the polarization caused by surface reaction is represented by a resistor $R_{in} = 2RT/Fai_0 L_R$ in each region. For galvanostatic discharging, $I_1$ and $I_2$ are subject to the constraint:

$$I_1 + I_2 = I \qquad \qquad 6)$$

where $I$ is the applied areal current density. Letting $\Phi_S$ at the current collector be $\Phi_S^0$ and setting $\Phi_L$ at the cathode/separator interface to be 0, the surface overpotentials in region I and II are given by

$$\eta_1 = U_{eq,1} - I_1 R_S - \Phi_S^0 \qquad \qquad 7)$$

$$\eta_2 = U_{eq,2} - I_2 R_L - \Phi_S^0 \qquad \qquad 8)$$

$U_{eq}$ is assumed to vary linearly with SOC or $c_s$: $U_{eq} = U_0 - \Delta U_{eq}(c_s - c_{s,0})/(c_{s,max} - c_{s,0})$, where $c_{s,0}$ and $c_{s,max}$ are the Li concentrations in the fully delithiated and lithiated states, respectively, and $U_0$ is $U_{eq}$ at SOC = 1. Accordingly, the evolution of $U_{eq}$ in region I and II during discharge is governed by the following equations:

$$\frac{dU_{eq,1}}{dt} = -\frac{2\Delta U_{eq}}{F(1-\epsilon_{cat})L_R(c_{S,max}-c_{S,0})} I_1 \equiv -\xi I_1 \qquad \qquad 9)$$

$$\frac{dU_{eq,2}}{dt} = -\xi I_2 \qquad \qquad 10)$$

where $\epsilon_{cat}$ is electrode porosity. Applying Eqs. 5 – 8 to eliminate $I_1$, $I_2$ and $\Phi_S^0$ in Eqs. 9 and 10, we obtain

$$\frac{dU_{eq,1}}{dt} = -\frac{\xi}{R_L+R_S+2R_{in}}\left[U_{eq,1} - U_{eq,2} + I(R_L + R_{in})\right] \qquad (11)$$

$$\frac{dU_{eq,2}}{dt} = -\frac{\xi}{R_L+R_S+2R_{in}}\left[U_{eq,2} - U_{eq,1} + I(R_S + R_{in})\right] \qquad (12)$$

from which $U_{eq,1}(t)$ and $U_{eq,2}(t)$ can be solved:

$$U_{eq,1}(t) = U_0 - \frac{I\xi t}{2} + \frac{I(R_S-R_L)}{4}\left(1 - \exp\left(-\frac{2\xi t}{R_L+R_S+2R_{in}}\right)\right) \qquad (13)$$

$$U_{eq,2}(t) = U_0 - \frac{I\xi t}{2} - \frac{I(R_S-R_L)}{4}\left(1 - \exp\left(-\frac{2\xi t}{R_L+R_S+2R_{in}}\right)\right) \qquad (14)$$

The equilibrium potential difference between region I and II, which quantifies the difference in the reaction degree, is thus

$$U_{eq,1} - U_{eq,2} = \Delta U_{ss}\left[1 - \exp\left(-\frac{t}{t_c}\right)\right] \qquad (15)$$

in which we define

$$\Delta U_{ss} \equiv \frac{I(R_S-R_L)}{2} = \frac{IL_R}{2}\left(\frac{1}{\sigma_{eff}} - \frac{1}{\kappa_{eff}}\right) \qquad (16)$$

$$t_c \equiv \frac{R_L+R_S+2R_{in}}{2\xi} = \frac{F(1-\epsilon)(c_{S,max}-c_{S,0})}{4\Delta U_{eq}}\left(\frac{L_R^2}{\kappa_{eff}} + \frac{L_R^2}{\sigma_{eff}} + \frac{4RT}{Fai_0}\right) \qquad (17)$$

The above result characterizes the time evolution of the reaction distribution within the reaction zone. As a main result of the circuit model, Eq. 15 shows that intercalation flux inside the reaction zone will reach a steady state distribution after a transient period with a characteristic time $t_c$. In the steady state, region I and II have an equal reaction current ($I_1 = I_2$) so that their equilibrium potential difference $U_{eq,1} - U_{eq,2}$ remains at a constant value $\Delta U_{ss}$. $\Delta U_{ss}$ is negative when discharge is electrolyte-transport-limited, or $\kappa_{eff} < \sigma_{eff}$, meaning that the active material near the separator will react first. For discharge limited by electronic transport ($\kappa_{eff} > \sigma_{eff}$), $\Delta U_{ss}$ is positive and the reaction will first occur near the current collector.

## III. Reaction Uniformity Number

In the last section, a steady-state equilibrium potential drop across the reaction zone, $\Delta U_{ss}$, is determined from the two-block circuit model. The fact that $|\Delta U_{ss}|$ cannot exceed $\Delta U_{eq}$ places an upper limit on $L_R$ that can be maintained during discharge, which is given by

$$L_{R,max} = \frac{2\Delta U_{eq}}{I\left|\frac{1}{\kappa_{eff}} - \frac{1}{\sigma_{eff}}\right|} \tag{18}$$

We introduce a dimensionless *reaction uniformity number* and defined it as the ratio between $L_{R,max}$ and the electrode thickness $L_{cat}$:

$$\lambda = \frac{L_{R,max}}{L_{cat}} = \frac{4\Delta U_{eq}}{IL_{cat}\left|\frac{1}{\kappa_{eff}} - \frac{1}{\sigma_{eff}}\right|} \tag{19}$$

$\lambda$ bears the physical meaning of the maximum normalized reaction zone width. Its magnitude provides a measure of the degree of the reaction uniformity within the porous electrode. When $\lambda \gg 1$, the entire electrode can establish a homogeneous reaction distribution and exhibit UR behavior. When $\lambda \ll 1$, reaction is confined to a narrow region much smaller than the electrode thickness, and so MZR-type behavior ensues. $\lambda$ reveals the roles of multiple factors (SOC dependence of $U_{eq}$, $\kappa_{eff}$, $\sigma_{eff}$, $L_{cat}$, $I$) in regulating the reaction distribution.

To examine the predicative power of $\lambda$ given by Eq. 19, we compare it against the P2D simulations of the model electrode system presented above in Figure 2. In Figure 4a, we plot the reaction zone width $W_{RZ}$, which is measured from the P2D simulations, against $\lambda$ estimated from the simulation parameters for electrodes with different $\Delta U_{eq}$. In evaluating $\lambda$, we use the electrolyte conductivity at $c = 1$ M for $\kappa_{eff}$ although concentration-dependent $\kappa_{eff}$ is used in simulations. It can be seen that the calculated $\lambda$ agrees very well with the measured $W_{RZ}$ over a wide range of values from 0.1 to 100, which shows that $\lambda$ can be used to accurately predict the extent of the reaction non-uniformity during discharge.

When the anion in the electrolyte has a non-zero transference number, local salt depletion (i.e. zero salt concentration) will occur in electrolyte near the current collector at high discharging rates or large electrode thickness(5, 11) and result in a large salt concentration gradient across the electrode, which will strongly influence the ionic conductivity. Although the simple circuit model presented in Section II does not take the concentration dependence of $k_{eff}$ into consideration, the reaction uniformity number $\lambda$ still provides a reliable indication of the electrode performance in the presence of salt depletion. In Figure 4b, we plot the normalized discharge capacity DoD$_f$ at 2C, 3C and 5C from the P2D simulations of the model electrode system as a function of $\Delta U_{eq}$. It shows that DoD$_f$ increases monotonically with $\Delta U_{eq}$ and has two plateaus at small and large $\Delta U_{eq}$ values, which correspond to the MZR and UR behavior, respectively. Previously, we developed a quantitative analytical model to predict the discharge performance of UR- and MZR-type electrodes in the electrolyte-diffusion-limited regime(5). It gives the expressions of the width of the salt penetration zone $L_{PZ}$ (i.e. region with non-zero salt concentration and complementary to the salt depletion zone) as listed in Table 1, and DoD$_f$ is evaluated as $L_{PZ}/L_{cat}$. The dashed and dash-dotted lines in Figure 4b represent DoD$_f$ predicted by the analytical model for MZR (DoD$_f^{MZR}$) and UR (DoD$_f^{UR}$) electrodes, respectively, which match the lower and upper limits of the simulated discharge capacity very well.

**Table 1**. Expressions of $L_{PZ}$ and $DoD_f$ for galvanostatic discharging of half cells predicted by an analytical model(5)

| $L_{PZ}$ | UR electrodes | $-\dfrac{3\epsilon_{sep}}{2\epsilon_{cat}}L_{sep} + \sqrt{\dfrac{6FD_{amb}c_0}{\tau_{cat}I(1-t_+)}(\epsilon_{cat}L_{cat} + \epsilon_{sep}L_{sep}) + \left(\dfrac{9\epsilon_{sep}^2}{4\epsilon_{cat}^2} - \dfrac{3\tau_{sep}}{\tau_{cat}}\right)L_{sep}^2}$ |
|---|---|---|
| | MZR electrodes | $-\dfrac{\epsilon_{sep}}{\epsilon_{cat}}L_{sep} + \sqrt{\dfrac{2FD_{amb}c_0}{\tau_{cat}I(1-t_+)}(\epsilon_{cat}L_{cat} + \epsilon_{sep}L_{sep}) + \left(\dfrac{\epsilon_{sep}^2}{\epsilon_{cat}^2} - \dfrac{\tau_{sep}}{\tau_{cat}}\right)L_{sep}^2}$ |

$$DoD_f = \begin{cases} L_{PZ}/L_{cat}, & 0 \leq L_{PZ} \leq L_{cat} \\ 1 & L_{PZ} > L_{cat} \end{cases}$$

While it is challenging to directly extend this analytical model to electrodes with the reaction behavior intermediate between UR and MZR, we find that $\lambda$ serves as a very good descriptor of $DoD_f$. When using Eq. 19 to calculate $\lambda$ in the presence of salt depletion, we replace $L_{cat}$ with $L_{PZ}$ for UR behavior, which is a more appropriate reference length scale as it represents the maximum thickness of the electrode region that can be fully discharged at the given discharge condition. In Figure 4c, we replot Figure 4b by rescaling $DoD_f$ as $\widetilde{DOD}_f = (DoD_f - DoD_f^{MZR})/(DoD_f^{UR} - DoD_f^{MZR})$, which always varies between 0 and 1. It clearly shows that the $\widetilde{DOD}_f \sim \lambda$ relations at different C rates collapse onto a single S-shaped curve, which can be fitted by an analytical function:

$$T(\lambda) = \frac{1}{2}[1 + \tanh(1.963 \log(\lambda) - 0.695)] \qquad 20)$$

To test the generality of Eq. 20, we performed 100 additional simulations by varying different cell parameters ($D_{amb}$, $L_{cat}$, $\epsilon_{cat}$, $\tau_{cat}$ and $L_{sep}$) and compared the simulated $\widetilde{DOD}_f$ against $T(\lambda)$. As shown in Figure 4d, overall the simulation results are in very good agreement with predictions by $T(\lambda)$, which likely represents a universal $\widetilde{DOD}_f \sim \lambda$ relation. The discharge capacity of electrodes

with intermediate reaction behavior in the electrolyte-transport-limited regime may therefore be predicted as

$$\text{DoD}_f = \text{DoD}_f^{\text{MZR}} + T(\lambda)\left(\text{DoD}_f^{\text{UR}} - \text{DoD}_f^{\text{MZR}}\right) \qquad 21)$$

**IV. Approaches to homogenizing reaction distribution in MZR-type electrodes**

Our work reveals that MZR-type electrodes, i.e. electrodes whose $U_{eq}$ is insensitive to SOC, have inferior performance at high rates and/or large electrode thickness due to the strong reaction inhomogeneity during discharge. In addition, the highly localized intercalation flux within the narrow reaction front may accelerate battery degradation by causing excessive stress concentration and local heat generation. Based on the insights from the P2D simulation and circuit model, we discuss in this section how reaction in this type of electrodes can be homogenized to make them more suitable for high rate and thick electrode applications. Somewhat counter-intuitively, we show that reducing the electronic conductivity and/or surface reaction rate is beneficial to improving the reaction uniformity in MZR-type electrodes.

**i) Reduce electronic conductivity**

The rate performance of today's Li-ion battery cells is typically limited by sluggish ionic transport in the electrolyte, whereas the electronic conductivity can be made sufficiently high with conductive additives or coatings on active materials. When $\kappa_{eff} \ll \sigma_{eff}$, electrode reaction first occurs near the separator upon discharging(12, 13), to which electrons travel a longer distance from the current collector to meet slow-moving Li ions from the anode. However, Eq. 19 predicts that the reaction uniformity can be improved by reducing the electronic conductivity to $\sigma_{eff} \approx \kappa_{eff}$ to render a large $\lambda$. To test this prediction, a P2D simulation is performed for a model system with $\Delta U_{eq} = 0.001\text{V}$, in which $\sigma_{eff}$ is set to $\kappa(c = 1M) = \kappa_0 \epsilon_{cat}^{1.5} = 0.291$ S/m. As shown in

Figure 5a, two reaction fronts form on both sides of the electrode and propagate towards the electrode center during 0.5C discharge. Accordingly, the intercalation flux is split into two peaks of lower intensities and does become more uniformly distributed compared to the higher $\sigma_{eff}$ case, see Figure 5b, although the reaction distribution is not entirely homogenized as predicted by the circuit model. This is because the model oversimplifies the situation by dividing the reaction zone into only two blocks and neglecting the non-uniformity within each block. Figure 5c shows that DoD$_f$ increases monotonically with decreasing $\sigma_{eff}$ upon discharging at higher rate (1 – 3C) and can even reach the discharge capacity of UR-type electrodes (dash-dotted lines) when $\sigma_{eff}$ approaches $\kappa_{eff}$.

**ii) Grade electronic conductivity**

Further improvement in the reaction uniformity can be realized by allowing $\sigma_{eff}$ to vary spatially within the electrode. This is because having a uniform reaction flux requires a constant surface overpotential everywhere, or

$$\nabla \eta = \nabla \Phi_L - \nabla \Phi_S + \nabla U_{eq} = 0 \qquad 22)$$

In Eq. 22, a significant $\nabla \Phi_L$ is usually present upon (dis)charging at relatively high rates due to the low ionic conductivity of the electrolyte, but $\nabla U_{eq} \approx 0$ for MZR-type materials. Replacing $\nabla \Phi_S$ and $\nabla \Phi_L$ with the current densities in the solid phase ($I_1$) and electrolyte ($I_2$), respectively, Eq. 22 becomes:

$$\frac{I_2}{\kappa_{eff}} - \frac{I_1}{\sigma_{eff}} = 0 \qquad 23)$$

In the presence of a uniform flux, both $I_1$ and $I_2$ vary linearly with the distance to the current collector $X$: $I_1 = I(L_{cat} - X)/L_{cat}$ and $I_2 = IX/L_{cat}$. Therefore, Eq. 23 is satisfied if the following relation between $\sigma_{eff}$ and $\kappa_{eff}$ holds:

$$\sigma_{eff} = \frac{L_{cat} - X}{X} \kappa_{eff} \qquad 24)$$

According to Eq. 24, the optimal $\sigma_{eff}$ is a hyperbolic function and varies monotonically from infinity at the current collector ($X = 0$) to 0 at the separator ($X = L_{cat}$). We confirm the effectiveness of such conductivity distribution via P2D simulation, in which $\sigma_{eff}$ is set as $\kappa_0 \epsilon^{1.5}(L_{cat} - X)/X$ and other parameters are the same as those for Figure 5a. As shown in Figure 5b and d, lithium intercalation indeed has a much more uniform distribution across the electrode throughout the discharge process than in systems with constant $\sigma_{eff}$.

We note that Palko et al.(14) recently describe a similar approach of tailoring spatially varied electrode matrix resistance to homogenize electrolyte depletion in electrical double layer capacitors (EDLC). The similarity in the derived $\sigma_{eff}$ expressions in ref. (14) and here demonstrates the analogy in the behavior of MZR-type battery electrodes and capacitors. On the other hand, UR-type electrodes behave in a very different way and Eq. 22 highlights such difference. The ability of UR-type compounds to sustain a non-zero spatial gradient in $U_{eq}$ in porous electrodes makes it possible to offset $\nabla \Phi_L$ with $\nabla U_{eq}$ to maintain a uniform overpotential without the need for spatially varied $\sigma_{eff}$. In the absence of a non-zero $\nabla U_{eq}$ in MZR-type electrodes, however, $\nabla \Phi_L$ can only be balanced by the $\Phi_s$ gradient to satisfy Eq. 22.

Experimentally, the electronic conductivity of porous electrodes can be tuned by adjusting the amount of conductive additives or applying coatings to active materials to either increase or decrease the conductivity, and graded electrodes may be prepared via layer-by-layer deposition processes. In Ref. (15), Zhang et al. fabricated layer-graded electrodes consisting of TiO$_2$(B) and reduced graphene oxide (RGO) and varied the RGO:TiO$_2$(B) ratio to control $\sigma_{eff}$ in each layer. They report that graded TiO$_2$(B)/RGO electrodes with the high $\sigma_{eff}$ layer placed adjacent to the current collector deliver more than 70% capacity at 20C than uniform electrodes with the same average RGO weight fraction. The theoretical analysis presented here explains why such an approach is effective. Using graded and heterogeneous architecture to enhance the rate performance of thick electrodes has been explored theoretically(16, 17) and experimentally(18-23) in recent years. Most existing studies focus on tailoring the porosity distribution to enhance electrolyte transport. Here we demonstrate a different strategy based on reducing the electronic conductivity to match the low ionic conductivity to improve the rate performance.

**iii) Reduce surface reaction rate**

The circuit model presented in Section II shows that during discharge a battery cell first goes through a transient period with a characteristic time $t_c$ before establishing a steady-state reaction zone within the porous electrode. Since the active material has a uniform SOC at the beginning of discharge, another way to improve the reaction uniformity in MZR-type electrodes is to increase $t_c$ to delay the establishment of the narrow reaction front and let the system remain in the transient period for the majority of the discharge process. Eq. 17 shows that $t_c$ can be increased by reducing $i_0$. We demonstrate this approach via P2D simulations of the model system with $\Delta U_{eq} = 0.001$ V, in which the surface reaction rate constant $k_0$ is decreased from the default value [$10^{-8}$ mol·m$^{-2}$·s$^{-1}$·(mol·m$^{-3}$)$^{-1.5}$] to $10^{-11}$ and $10^{-13}$ mol·m$^{-2}$·s$^{-1}$·(mol·m$^{-3}$)$^{-1.5}$. The corresponding time evolution of $\tilde{c}_S(\tilde{X})$ is plotted in Figure 6a and 6b, respectively. Compared to Figure 3b, a smaller $k_0$ indeed slows down the development of the sharp reaction front and results in a more uniform reaction across the electrode during discharge. As expected, decreasing $k_0$ can increase the discharge capacity by 20 – 35% at high rates (1 – 3C), Figure 6c, which is similar to the effect of reducing $\sigma_{eff}$ although the improvement is not as pronounced. The reason that reducing the surface reaction kinetics is beneficial is that it prevents the localization of the intercalation flux and forces the reaction current to spread out over a larger electrode region. Experimentally, surface reaction kinetics may be tailored by "artificial" SEI such as ALD coatings of various inorganic compounds (e.g. $Al_2O_3$, $TiO_2$, $ZrO_2$ (24-26)), whose insulating nature could retard the intercalation process and cause higher surface polarization.

While reducing $\sigma_{eff}$ and $k_0$ promotes the reaction uniformity, such approaches may lead to increased energy loss and degrade energy efficiency, the severity of which needs to be examined. Figure 7a shows the cell potential curves from the simulations of discharging four types of 200μm-thick model electrodes ($\Delta U_{eq}$= 0.001 V) at 1C: high $\sigma_{eff}$ (100 S/m, baseline), low and uniform $\sigma_{eff}$ (0.291 S/m), graded $\sigma_{eff}$ (Eq. 24), and low surface reaction rate $k_0$ [$10^{-13}$ mol·m$^{-2}$·s$^{-1}$·(mol·m$^{-3}$)$^{-1.5}$]. The electrodes with low and graded $\sigma_{eff}$ only see a small drop (~0.02 V) in the discharge potential while delivering 30% more capacity than the baseline electrode at the same time. On the other hand, the electrode with reduced $k_0$ has a larger depression in the discharge potential (~0.1 V). In Figure 7b, the total energy loss in a half cell upon 1C discharging, which is the sum of losses due to the ionic, electronic and surface reaction resistances, is plotted as a function of DoD for the four cases. It can be seen that reducing or grading $\sigma_{eff}$ only slightly increases the energy loss by less than 15% compared to the baseline case while decreasing $k_0$ doubles the energy loss. Therefore, tailoring the electronic conductivity of the solid electrode phase is a more attractive strategy to enhance the discharge performance of thick electrodes.

**Conclusion**

In this work, we employ P2D simulations and equivalent circuit model to elucidate the important role of the SOC dependence of the open-circuit potential $U_{eq}$, an intrinsic thermodynamic property of battery compounds, in controlling the reaction uniformity within porous electrodes. Electrode reaction becomes increasingly homogeneous with the slope of the $U_{eq}$(SOC) curve, which has a direct impact on the battery discharge performance at high rates. The limiting cases can be described by the "uniform reaction" or UR behavior for electrodes whose $U_{eq}$ has strong SOC dependence (e.g. NMC and NCA), and the "moving-zone reaction" or MZR behavior for electrodes with SOC-independent $U_{eq}$ (e.g. LiFePO$_4$, Li$_4$Ti$_5$O$_{12}$). A dimensionless "reaction uniformity" number, $\lambda = 4\Delta U_{eq} \left| \kappa_{eff}^{-1} - \sigma_{eff}^{-1} \right|^{-1} / IL_{cat}$, is introduced to capture the effects of electrode and cycling parameters on the degree of reaction inhomogeneity. In the electrolyte-transport-limited regime, $\lambda$ accurately predicts the reaction zone width and exhibits a universal correlation with the rescaled discharge capacity, making $\lambda$ a useful indicator of the electrode performance. We show that the reaction distribution in MZR-type electrodes can be homogenized by several approaches including 1) matching the ionic and electronic conductivities, 2) grading the electronic conductivity, and 3) slowing down the surface reaction kinetics, of which the first two do not significantly reduce the energy efficiency of the discharging process.

## Appendix A

P2D simulations: detailed description of the P2D model can be found in literature(6-10). The governing equations implemented in the half cell simulations are summarized as follows.

The concentration $c$ and ionic current $\mathbf{i}$ in a binary electrolyte are given by

$$\epsilon_i \frac{\partial c}{\partial t} = \nabla \cdot \left( \frac{\epsilon_i}{\tau_i} D_{amb} \nabla c \right) + \nabla \cdot \left( \frac{(1-t_+)\mathbf{i}}{F} \right) \qquad \text{A1)}$$

$$\mathbf{i} = -\frac{\epsilon_i}{\tau_i} \kappa(c) \nabla \Phi_L - \frac{\epsilon_i}{\tau_i} \frac{RT\kappa(c)(2t_+ - 1)}{Fc} \left( 1 + \frac{\partial \ln f_\pm}{\partial \ln c} \right) \nabla c \qquad \text{A2)}$$

where the subscript $i = cat$ or $sep$ to represent cathode or separator. Let $X = 0$ be at the interface between the current collector and cathode. The boundary conditions are

$$\left. \frac{\partial c}{\partial X} \right|_{X=0} = 0 \qquad \text{A3.1)}$$

$$\left. \frac{\epsilon_{cat}}{\tau_{cat}} D_{amb} \frac{\partial c}{\partial X} \right|_{X=L_{cat}^-} = \left. \frac{\epsilon_{sep}}{\tau_{sep}} D_{amb} \frac{\partial c}{\partial X} \right|_{X=L_{cat}^+} \qquad \text{A3.2)}$$

$$\left. \frac{\epsilon_{sep}}{\tau_{sep}} D_{amb} \frac{\partial c}{\partial X} \right|_{X=L_{cat}+L_{sep}} = \frac{I(1-t_+)}{F} \qquad \text{A3.3)}$$

$$\mathbf{i}|_{X=0} = 0 \qquad \text{A4.1)}$$

$$\mathbf{i}|_{X=L_{cat}^-} = \mathbf{i}|_{X=L_{cat}^+} \qquad \text{A4.2)}$$

$$\mathbf{i}|_{X=L_{cat}+L_{sep}} = i_0^{Li} \left[ \exp\left( \frac{\alpha F(\Phi_L - \Phi_{Li})}{RT} \right) - \exp\left( -\frac{(1-\alpha)F(\Phi_L - \Phi_{Li})}{RT} \right) \right] = -I \qquad \text{A4.3)}$$

where superscripts (+ and -) denote the right and left side of the cathode/separator interface, respectively, and the potential of Li anode $\Phi_{Li}$ is fixed at 0. The electronic current in solid electrode matrix $\mathbf{i_S}$ is

$$\mathbf{i_S} = -\sigma_{eff} \nabla \Phi_S \tag{A5}$$

with the boundary conditions

$$\mathbf{i_S}|_{X=0} = -I \tag{A6.1}$$

$$\mathbf{i_S}|_{X=L_{cat}} = 0 \tag{A6.2}$$

The ionic and electronic currents are coupled by surface reaction as

$$\nabla \cdot \mathbf{i} = -\nabla \cdot \mathbf{i_S} = -Faj_{in} \tag{A7}$$

where $a = 3(1 - \epsilon_{cat})/r_{cat}$ in cathode and reaction flux density $j_{in}$ follows the Butler-Volmer equation (Eq. 1), in which $i_0$ is given by

$$i_0 = Fk_0 c^{1-\alpha} c_S^\alpha (c_{S,max} - c_S)^\alpha \tag{A8}$$

In the P2D simulations of NMC and LFP half cells, lithium diffusion in active materials is simplified as a radial diffusion in spherical particles as

$$\frac{\partial c_S}{\partial t} = \frac{1}{r^2} \frac{\partial}{\partial r}\left(r^2 D_S \frac{\partial c_S}{\partial r}\right) \tag{A9}$$

where the boundary conditions are

$$\left.\frac{\partial c_S}{\partial r}\right|_{r=0} = 0 \tag{A10.1}$$

$$D_S \left(\frac{\partial c_S}{\partial r}\right)\bigg|_{r=r_{cat}} = j_{in} \qquad \text{A10.2)}$$

For the model electrode material, Li diffusion is assumed to be very facile so that $c_s$ is constant within each electrode particle. All of the simulations are implemented in COMSOL Multiphysics® 5.3a.

**Symbol list**

| | |
|---|---|
| $a$ | Volumetric surface area of cathode [m$^{-1}$] |
| $C$ | C rate |
| $c$ | Salt concentration in electrolyte [mol·m$^{-3}$] |
| $c_0$ | Initial salt concentration in electrolyte [mol·m$^{-3}$] |
| $c_S$ | Li concentration in electrode particles [mol·m$^{-3}$] |
| $\tilde{c}_S$ | Normalized Li concentration in electrode particles |
| $c_{S,0}$ | Initial Li concentration in electrode particles [mol·m$^{-3}$] |
| $c_{S,max}$ | Maximum Li concentration in electrode particles [mol·m$^{-3}$] |
| $D_{amb}$ | Ambipolar diffusivity of electrolyte [m$^2$·s$^{-1}$] |
| $D_s$ | Li diffusivity in active material [m$^2$·s$^{-1}$] |
| DoD | Depth of discharge |
| DoD$_f$ | Final depth of discharge or normalized discharge capacity |
| $\widetilde{\text{DoD}_f}$ | Rescaled final depth of discharge |
| DOD$_f^{MZR}$ / DOD$_f^{UR}$ | Predicted final depth of discharge of moving-zone reaction / uniform reaction |
| $F$ | Faraday constant (96485 C·mol$^{-1}$) |
| $f_{Li}$ | Li fraction in active material |
| $I$ | Applied current density [A·m$^{-2}$] |
| $I_1$ / $I_2$ | Current density in solid / liquid phase [A·m$^{-2}$] |
| $i_0$ | Exchange current density of active material [A·m$^{-2}$] |
| $i_0^{Li}$ | Exchange current density on Li anode [A·m$^{-2}$] |
| $j_{in}$ | Reaction flux on active material surface [mol·m$^{-2}$·s$^{-1}$] |
| $k_0$ | Reaction rate constant [mol·m$^{-2}$·s$^{-1}$·(mol·m$^{-3}$)$^{-1.5}$] |
| $L_{cat}$ / $L_{sep}$ | Cathode / separator thickness [m] |
| $L_{PZ}$ | Salt penetration depth [m] |
| $L_R$ | Reaction zone length in the circuit model [m] |
| $n$ | Number of electrons in equation for electrode reaction |
| $R$ | Gas constant (8.314 J·mol$^{-1}$·K$^{-1}$) |
| $R_L$ / $R_S$ / $R_{in}$ | Resistance in liquid phase / in solid phase / on particle surface [Ω/m$^2$] |
| $r_{cat}$ | Cathode particle radius [m] |
| SOC | State of charge |
| $T$ | Temperature [298 K] |
| $t_c$ | Characteristic time before reaching steady state [s] |
| $t_+$ | Cation transference number in electrolyte |
| $U_{eq}$ | Equilibrium (open-circuit) potential of active material [V] |
| $\Delta U_{eq}$ | Slope of equilibrium potential [V] |
| $\Delta U_{SS}$ | Potential difference between two regions at steady state [V] |
| $W_{RZ}$ | Scaled reaction zone width measured from P2D simulation |
| $X$ | Spatial coordinate [m] |
| $\tilde{X}$ | Spatial coordinate normalized by $L_{cat}$ |
| $1+\partial \ln f_\pm / \partial \ln c$ | Thermodynamic factor |
| $\alpha$ | Charge transfer coefficient |
| $\epsilon_{cat} / \epsilon_{sep}$ | Cathode / separator porosity |
| $\eta$ | Overpotential [V] |
| $\kappa_{eff}$ | Effective electrolyte conductivity [S·m$^{-1}$] |
| $\kappa_0$ | Reference electrolyte conductivity at 1M [S·m$^{-1}$] |
| $\lambda$ | Reaction uniformity number |
| $\sigma_{eff}$ | Effective solid phase conductivity [S·m$^{-1}$] |
| $\tau_{cat} / \tau_{sep}$ | Cathode / separator tortuosity |
| $\Phi_L / \Phi_S$ | Electrolyte / solid phase potential [V] |
| $\Phi_S^0$ | Solid phase potential near current collector [V] |

Table A1. Parameters used in P2D simulations (unless otherwise stated)

| Parameter | Symbol | Value | | |
|---|---|---|---|---|
| Electrode properties | | | | |
| | | NMC | LFP | Model cathode material |
| Cathode particle radius (μm) | $r_{cat}$ | 1 | 0.1 | 0.1 |
| Cathode porosity | $\epsilon_{cat}$ | 0.25 | 0.25 | 0.25 |
| Separator thickness (μm) | $L_{sep}$ | 25 | | |
| Separator porosity | $\epsilon_{sep}$ | 0.55 | | |
| Tortuosity | $\tau$ | $\tau = \epsilon^{-0.5}$ | | |
| Maximum Li concentration in active materials (mol·m$^{-3}$) | $c_{S,max}$ | 49761 | 22806 | 20000 |
| Initial concentration in active materials (mol·m$^{-3}$) | $c_{S,0}$ | 22392 | 228 | 200 |
| Li diffusivity in active materials (m$^2$·s$^{-1}$) | $D_S$ | 10$^{-14}$ (27) | 10$^{-16}$ (28) | -- |
| Effective electrode conductivity (S·m$^{-1}$) | $\sigma_{eff}$ | 10 (29) | 10 [a] | 100 |
| Reaction rate constant (mol·m$^{-2}$·s$^{-1}$·(mol·m$^{-3}$)$^{-1.5}$) | $k_0$ | 3·10$^{-11}$ (29) | 3·10$^{-11}$ [a] | 10$^{-8}$ |
| Charge transfer coefficient | $\alpha$ | 0.5 | | |
| Exchange current density of Li anode (A·m$^{-2}$) | $i_0^{Li}$ | 20 (29) | | |
| Equilibrium potential (V) | $U_{eq}$ | See note b | | |
| Electrolyte (1M LiPF$_6$ in EC/DMC 50:50 wt.%) properties | | | | |
| Initial salt concentration (mol·m$^{-3}$) | $c_0$ | 1000 | | |
| Transference number of cations | $t_+$ | 0.39 (30) | | |
| Ambipolar diffusivity (m$^2$·s$^{-1}$) | $D_{amb}$ | 2.95·10$^{-10}$ (30) | | |
| Concentration-dependent ionic conductivity (S·m$^{-1}$) | $\kappa(c)$ | $0.00233c$ [c] | | |

Note:
  a. Assumed.
  b. The equilibrium potential profiles of NMC and LFP are extracted from Figure 2 in Ref. (31), and Figure 2 in Ref. (32). The equilibrium potential of model cathode material is defined by Eq. 4.
  c. Calculated by $\kappa = F^2 D_{amb} c / \left( 2RT t_+ (1-t_+) \right)$

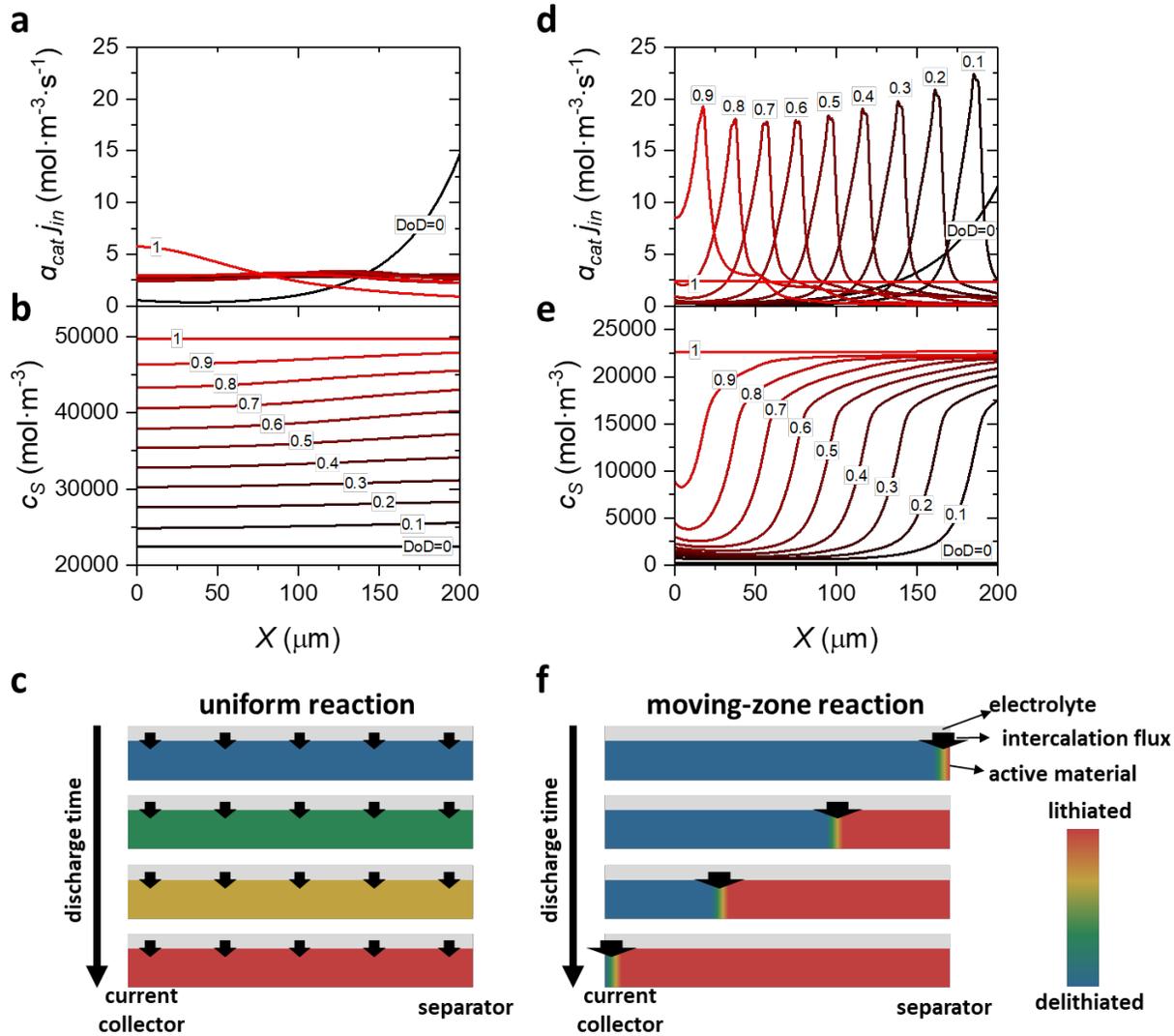

**Figure 1**. UR / MZR reaction behavior displayed by NMC111 / LFP half cells. In P2D simulations, both cells have a cathode thickness $L_{cat}$ = 200 μm and are discharged at $I$=5mA/cm$^2$. Other simulation parameters are listed in Table A1. **a** and **d**. Reaction flux $a_{cat}j_{in}$ on NMC111 and LFP particle surface, respectively. **b** and **d**. Average Li concentration $c_S$ in NMC111 and LFP particles, respectively. **c** and **f**. Schematics of idealized UR vs MZR behavior.

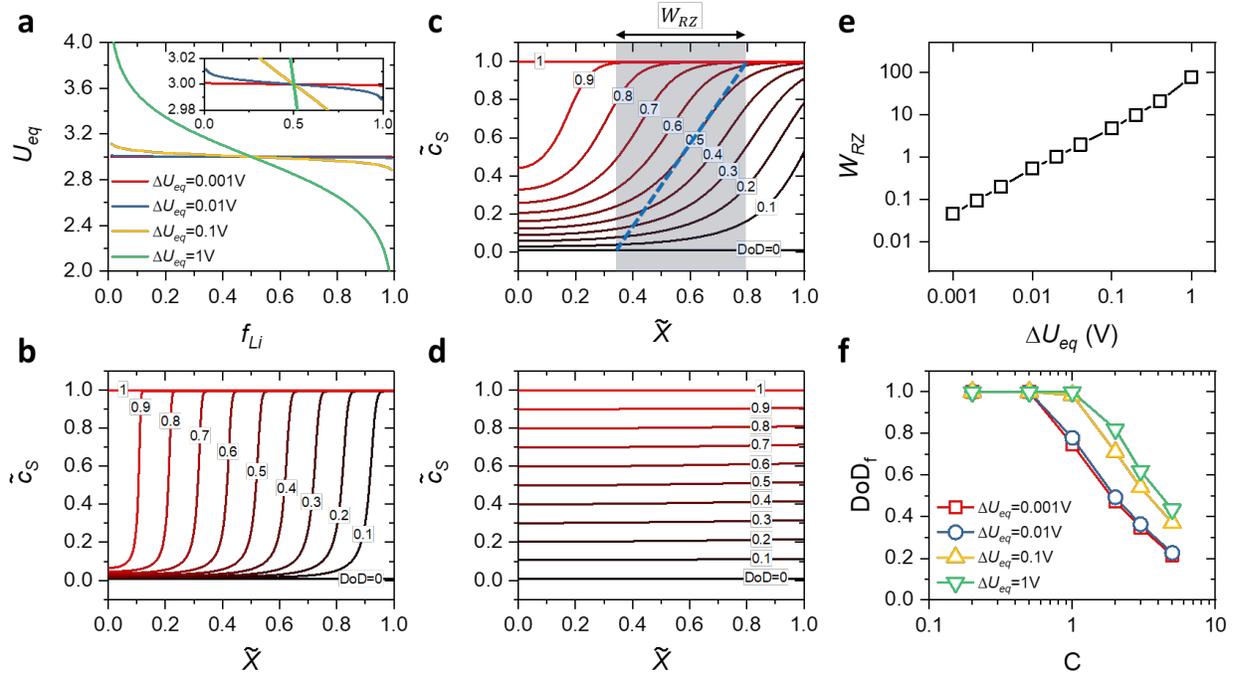

**Figure 2**. P2D simulations of a model electrode material in half cell configuration. $L_{cat}$=200μm and other simulation parameters are listed in Table A1. **a** $U_{eq} - f_{Li}$ relation of the model electrode with different slope $\Delta U_{eq}$ at $f_{Li}$ = 0.5. **b-d** Spatial distribution of $\tilde{c}_S$ at different DoDs upon 0.5C discharging for electrodes with $\Delta U_{eq}$ = 0.001V (**b**), 0.01V (**c**) and 1 V (**d**). $\tilde{X}$ is the scaled distance to the current collector (separator at $\tilde{X}$ = 1). The dashed line and shaded area in **c** illustrate the scaled reaction zone width $W_{RZ}$ defined in the text. **e**. Dependence of $W_{RZ}$ on $\Delta U_{eq}$. **f**. Normalized discharge capacity vs C rate in half cells for electrodes with $\Delta U_{eq}$ = 0.001, 0.01, 0.1 and 1 V.

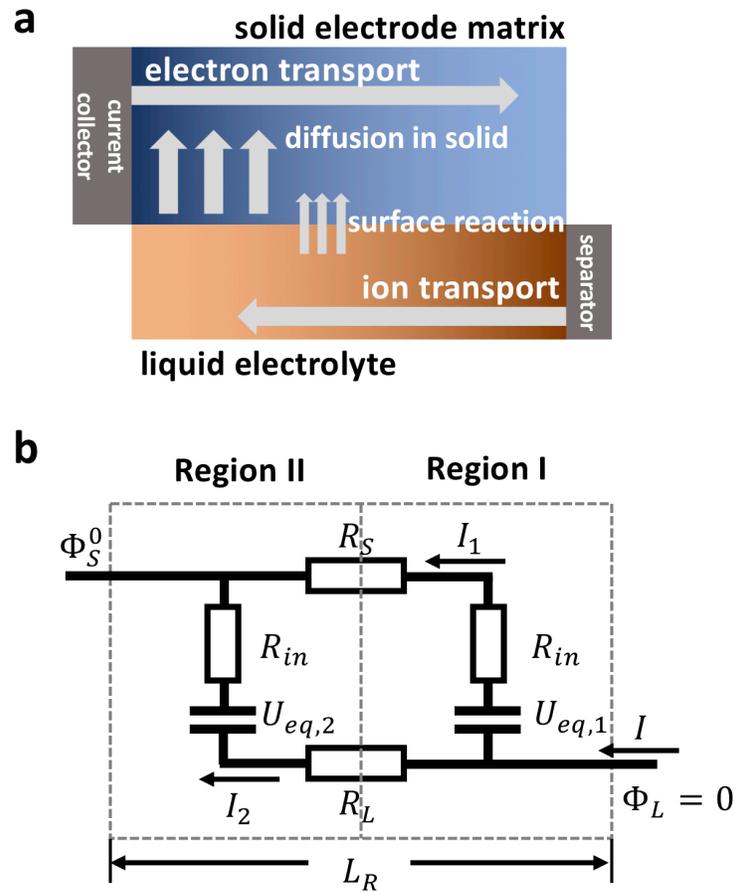

**Figure 3.** **a** Schematic of relevant kinetic processes during discharge and **b** their representation in a two-block equivalent circuit model.

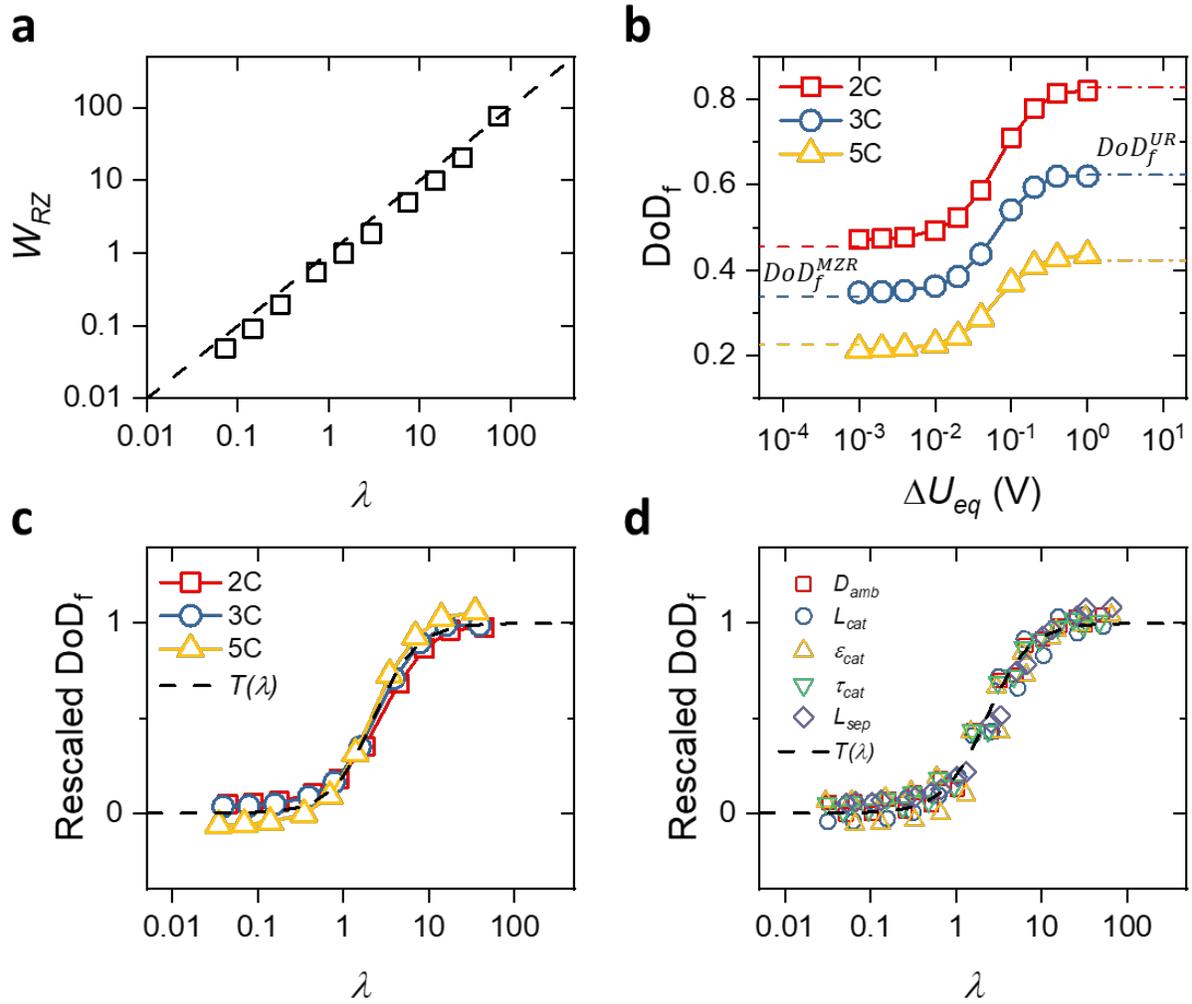

**Figure 4**. **a** Comparison between $W_{RZ}$ in Figure 2e and $\lambda$ predicted by Eq. 19. Dashed line corresponds to $W_{RZ} = \lambda$. **b** Normalized discharge capacity $DoD_f$ vs $\Delta U_{eq}$ at 2C, 3C and 5C. Dashed and dash-dotted lines are the predicted $DoD_f$ for MZR-type electrodes ($DoD_f^{MZR}$) and UR-type electrodes ($DoD_f^{UR}$) based on the analytical model listed in Table 1, respectively. **c.** Rescaled discharge capacity $\widetilde{DoD_f}$ ($\frac{DoD_f - DoD_f^{MZR}}{DoD_f^{UR} - DoD_f^{MZR}}$) vs $\lambda$ at 2C, 3C and 5C. **d**. Test of the sensitivity of the $\widetilde{DoD_f} \sim \lambda$ relation to simulation parameters. $D_{amb}$, $L_{cat}$, $\epsilon_{cat}$ and $\tau_{cat}$ are individually varied to 0.7× and 1.3 × of their default values ($D_{amb}$ = 2.95·10⁻¹⁰ m²/s, $L\_cat$ = 200 μm, $\epsilon_{cat}$ = 0.25, $\tau_{cat}$ = 2), and $L_{sep}$ is set to 2× and 3 × of its default value (25 μm). Discharge rate is fixed at 3C and other parameters are the same as in Table A1. The black dashed line in **c** and **d** is the transition function $T(\lambda)$ (Eq. 20).

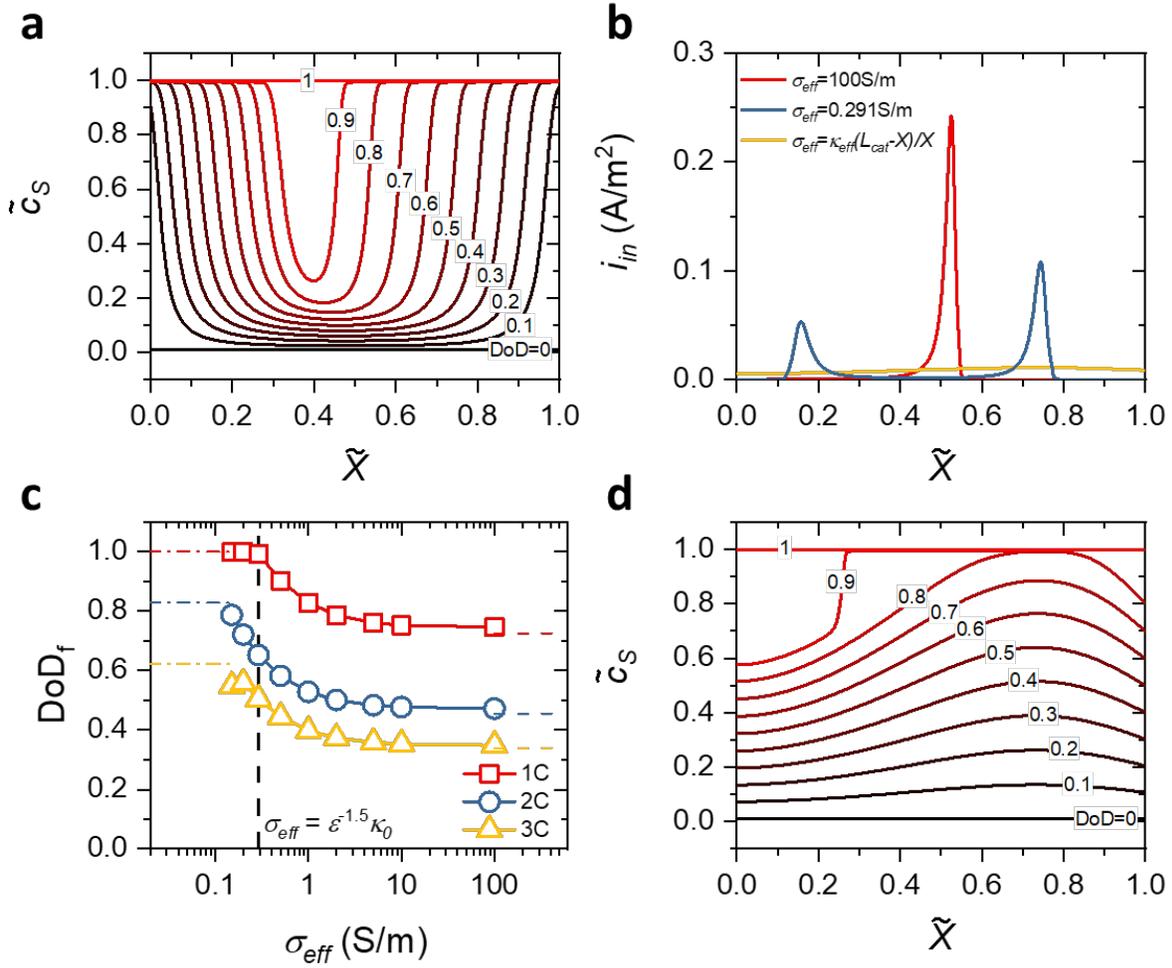

**Figure 5**. **a**. Spatial distribution of $\tilde{c}_S$ at different DoDs in a 200μm-thick model electrode in a half cell discharged at 0.5C with $\Delta U_{eq} = 0.001$ V and $\sigma_{eff} = 0.291$ S/m. **b**. Reaction current distributions at DoD = 0.5 for electrodes with $\sigma_{eff} = 100$ S/m, 0.291 S/m and $\kappa_0 \epsilon^{1.5}(L_{cat} - X)/X$, where $X$ is the distance to the current collector. **c**. Effect of $\sigma_{eff}$ on the discharge capacity at 1C, 2C and 3C discharging. Dashed and dash-dotted lines are $\text{DoD}_f^{\text{MZR}}$ and $\text{DoD}_f^{\text{UR}}$ predicted by the analytical model, respectively. **d**. Spatial distribution of $\tilde{c}_S$ at different DoDs in a 200μm-thick model electrode with $\Delta U_{eq} = 0.001$ V and variable electronic conductivity $\sigma_{eff} = \kappa_0 \epsilon^{1.5}(L_{cat} - X)/X$ upon 0.5C discharging.

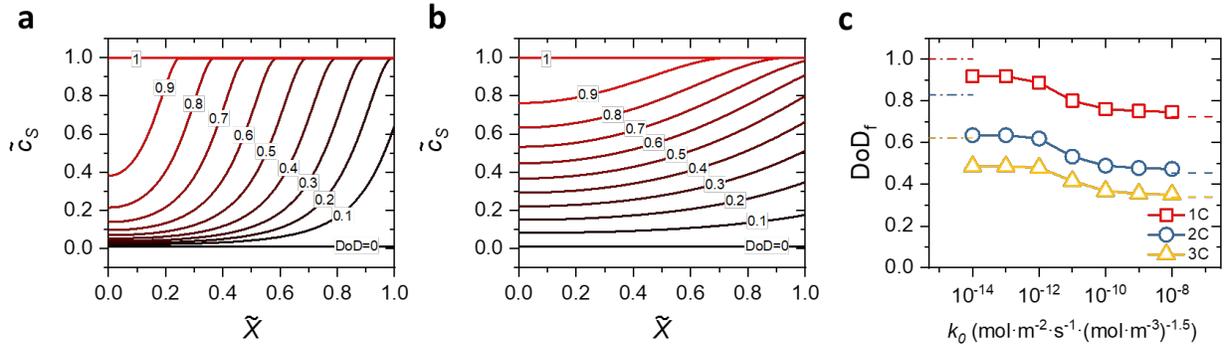

**Figure 6.** P2D simulations of discharging a 200μm-thick model electrode ($\Delta U_{eq}$= 0.001 V) with low surface reaction rate constant $k_0$ in a half cell at 1C. Other simulation parameters are the same as those for Figure 3b. **a** and **b.** Spatial distribution of $\tilde{c}_S$ at different DoD during 0.5C discharging with $k_0=10^{-11}$ (**a**) and $k_0=10^{-13}$ (**b**) mol·m$^{-2}$·s$^{-1}$·(mol·m$^{-3}$)$^{-1.5}$. **c.** Discharge capacity vs $k_0$ upon 1C, 2C and 3C discharging. Dashed and dash-dotted lines are $DoD_f^{MZR}$ and $DoD_f^{UR}$ predicted by the analytical model (Table 1).

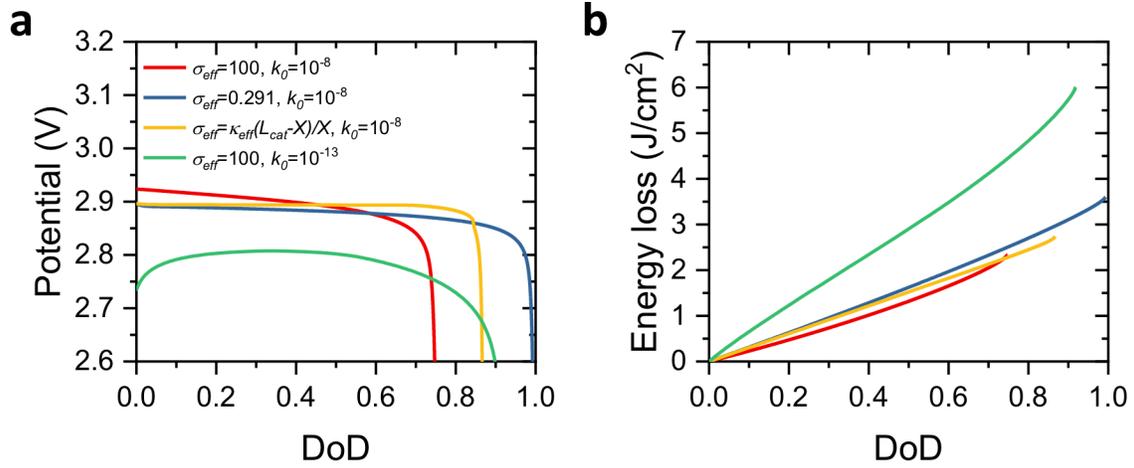

**Figure 7 a**. Discharge potential curves of 200μm-thick model electrodes ($\Delta U_{eq}$= 0.001 V) with different electronic conductivity $\sigma_{eff}$ or surface reaction rate constant $k_0$ values as shown in the legend. Unit of $\sigma_{eff}$ is S/m and of $k_0$ is mol·m$^{-2}$·s$^{-1}$·(mol·m$^{-3}$)$^{-1.5}$. Electrodes are discharged at 1C in half cells. **b**. Energy loss at the cell level vs DoD.

# References

1. J. Euler and W. Nonnenmacher, *Electrochimica Acta*, **2**, 268 (1960).
2. O. S. Ksenzhek and V. V. Stender, *Doklady Akad. Nauk S.S.S.R.*, **106**, 487 (1956).
3. O. S. Ksenzhek and V. V. Stender, *Doklady Akad. Nauk S.S.S.R.*, **107**, 280 (1956).
4. J. S. Newman and C. W. Tobias, *Journal of the Electrochemical Society*, **109**, 1183 (1962).
5. F. Wang and M. Tang, *under review (https://arxiv.org/abs/2004.10707)* (2020).
6. M. Doyle, T. F. Fuller and J. Newman, *Journal of the Electrochemical Society*, **140**, 1526 (1993).
7. T. F. Fuller, M. Doyle and J. Newman, *Journal of the Electrochemical Society*, **141**, 1 (1994).
8. V. Srinivasan and J. Newman, *Journal of The Electrochemical Society*, **151**, A1530 (2004).
9. K. E. Thomas, J. Newman and R. M. Darling, in *Advances in Lithium-Ion Batteries*, edited by W. van Schalkwijk and B. Scrosati, Kluwer Academic/Plenum Publishers, New York (2002).
10. T. R. Ferguson and M. Z. Bazant, *Journal of the Electrochemical Society*, **159**, A1967 (2012).
11. K. G. Gallagher, S. E. Trask, C. Bauer, T. Woehrle, S. F. Lux, M. Tschech, P. Lamp, B. J. Polzin, S. Ha, B. Long, Q. L. Wu, W. Q. Lu, D. W. Dees and A. N. Jansen, *Journal of the Electrochemical Society*, **163**, A138 (2016).
12. T. Sasaki, C. Villevieille, Y. Takeuchi and P. Novak, *Advanced Science*, **2**, 1500083 (2015).
13. J. Liu, M. Kunz, K. Chen, N. Tamura and T. J. Richardson, *The Journal of Physical Chemistry Letters*, **1**, 2120 (2010).
14. J. W. Palko, A. Hemmatifar and J. G. Santiago, *Journal of Power Sources*, **397**, 252 (2018).
15. Y. Y. Zhang, O. I. Malyi, Y. X. Tang, J. Q. Wei, Z. Q. Zhu, H. R. Xia, W. L. Li, J. Guo, X. R. Zhou, Z. Chen, C. Persson and X. D. Chen, *Angew Chem Int Edit*, **56**, 14847 (2017).
16. Y. Dai and V. Srinivasan, *Journal of The Electrochemical Society*, **163**, A406 (2015).
17. Y. Qi, T. Jang, V. Ramadesigan, D. T. Schwartz and V. R. Subramanian, *Journal of The Electrochemical Society*, **164**, A3196 (2017).
18. L. Liu, P. J. Guan and C. H. Liu, *Journal of the Electrochemical Society*, **164**, A3163 (2017).
19. C. J. Bae, C. K. Erdonmez, J. W. Halloran and Y. M. Chiang, *Advanced Materials*, **25**, 1254 (2013).
20. J. S. Sander, R. M. Erb, L. Li, A. Gurijala and Y. M. Chiang, *Nature Energy*, **1** (2016).
21. L. S. Li, R. M. Erb, J. J. Wang, J. Wang and Y. M. Chiang, *Advanced Energy Materials*, **9** (2019).
22. J. Billaud, F. Bouville, T. Magrini, C. Villevieille and A. R. Studart, *Nature Energy*, **1** (2016).
23. B. Delattre, R. Amin, J. Sander, J. De Coninck, A. P. Tomsia and Y. M. Chiang, *Journal of the Electrochemical Society*, **165**, A388 (2018).
24. Y. S. Jung, A. S. Cavanagh, L. A. Riley, S. H. Kang, A. C. Dillon, M. D. Groner, S. M. George and S. H. Lee, *Advanced Materials*, **22**, 2172 (2010).
25. X. F. Li, J. Liu, X. B. Meng, Y. J. Tang, M. N. Banis, J. L. Yang, Y. H. Hu, R. Y. Li, M. Cai and X. L. Sun, *Journal of Power Sources*, **247**, 57 (2014).
26. J. C. Ye, Y. H. An, T. W. Heo, M. M. Biener, R. J. Nikolic, M. Tang, H. Jiang and Y. M. Wang, *Journal of Power Sources*, **248**, 447 (2014).
27. R. Amin and Y. M. Chiang, *Journal of the Electrochemical Society*, **163**, A1512 (2016).
28. B. Rajabloo, A. Jokar, M. Desilets and M. Lacroix, *Journal of the Electrochemical Society*, **164**, A99 (2017).
29. Z. Mao, M. Farkhondeh, M. Pritzker, M. Fowler and Z. Chen, *Journal of the Electrochemical Society*, **163**, A458 (2016).
30. P. Porion, Y. R. Dougassa, C. Tessier, L. El Ouatani, J. Jacquemin and M. Anouti, *Electrochimica Acta*, **114**, 95 (2013).
31. J. Smekens, J. Paulsen, W. Yang, N. Omar, J. Deconinck, A. Hubin and J. Van Mierlo, *Electrochimica Acta*, **174**, 615 (2015).
32. M. Wang, J. J. Li, X. M. He, H. Wu and C. R. Wan, *Journal of Power Sources*, **207**, 127 (2012).